\documentclass[%
 reprint,
 amsmath,amssymb,
 aps,
]{revtex4-2}

\usepackage{graphicx}% Include figure files
\usepackage{dcolumn}% Align table columns on decimal point
\usepackage{bm}% bold math
\usepackage[notrig]{physics}
\begin{document}

\preprint{APS/123-QED}

\title{Suppression of Richtmyer-Meshkov instability via special pairs of shocks and phase transitions}% Force line breaks with \\
%W.J.~Schill${}^1$, H.~Lorenzana,..., J.~Belof
\author{W. J.~Schill}
\email{schill1@llnl.gov}
 %\altaffiliation[Also at ]{Physics Department, XYZ University.}%Lines break automatically or can be forced with \\J.~Belof
 
 %Mike Armstrong, Jeff Nguyen, Dan White, Lorin Benedict, Brandon La Lone, Mike Staska, Andrew Hoff, Dane Sterbentz

 \author{M. R.~Armstrong}%
 \author{J. H.~Nguyen}%
 \author{D. M.~Sterbentz}%
 \author{D. A.~White}%
 \author{L. X.~Benedict}%
 \author{R. N.~Rieben}%
 \author{A.~Hoff}% 
\author{H. E.~Lorenzana}%
\author{J. L.~Belof}%
\affiliation{%
 Lawrence Livermore National Laboratory,
 7000 East ave, Livermore, CA 94550, USA. \\
 %This line break forced with 
}%


\author{B. M. La Lone}
\author{ M. D. Staska}
%\affiliation{
\affiliation{
 Special Technologies Laboratory
 5520 Ekwill St b, Santa Barbara, CA 93117
}%
%\affiliation{%


\date{\today}% It is always \today, today,

\begin{abstract}
 The classical Richtmyer-Meshkov instability is a hydrodynamic instability characterizing the evolution of an interface following shock loading. In contrast to other hydrodynamic instabilities such as Rayleigh-Taylor, it is known for being unconditionally unstable: regardless of the direction of shock passage, any deviations from a flat interface will be amplified. In this article, we show that for negative Atwood numbers, there exist special sequences of shocks which result in a nearly perfectly suppressed instability growth. We demonstrate this principle computationally and experimentally with stepped fliers and phase transition materials. A fascinating immediate corollary is that in specific instances a phase transitioning material may self-suppress RMI.  
\end{abstract}

\maketitle

\section{\label{sec:level1} Introduction }

 The classical Richtmyer-Meshkov instability (RMI) is a hydrodynamic instability characterizing the evolution of an interface following shock loading. 
 In the case of a shock passing from a heavy material to a light material, the evolution of the interface follows a standard behavior, valleys evolve into peaks or \textit{jets}  and the initial peaks evolve into valleys.
 In contrast to other hydrodynamic instabilities such as Rayleigh Taylor (RT), RMI is known for being unconditionally unstable: regardless of the direction of shock passage, any deviations from a flat interface will be amplified. 
 RMI has held a critical role in both scientific and technological applications including astrophysics \cite{zhou2021rayleigh}, mining \cite{birkhoff1948explosives}, many applications of fluid transport \cite{zhou2019turbulent} including scramjets \cite{zhou2021rayleigh}, and laser driven inertial confinement fusion (ICF) \cite{mikaelian2011extended,remington2019rayleigh} such as is pursued at the National Ignition Facility (NIF). 
 For ICF, in particular, RMI is considered a critical limiting physical mechanism owing to its role in onset of mix which ultimately may quench burn \cite{dimonte2013ejecta,remington2019rayleigh}; in a certain sense, the degradation due to RMI is one of the key bottlenecks to development of abundant clean energy via fusion.
 For these reasons, it is of great importance to develop methodologies to suppress growth of RMI.
 %
 %

 %
 %In this article, we show that for negative Atwood numbers, there exist special sequences of shocks which result in a (nearly) perfectly suppressed instability. 
 
 RMI has received intense study; see the reviews \cite{ZHOU20171,zhou2017rayleigh} and the references therein. 
 A productive description of RMI in terms of the vorticity field (cf. \cite{chorin1994vorticity}) was introduced in \cite{hawley1989vortex,jacobs1996experimental,likhachev2005vortex}. See also \cite{sterbentz2022design}.
 In this conception of the instability physics, the passage of the shock through the interface deposits vorticity at the interface via the baroclinic mechanism. 
 The variation in the sign of the vorticity arising from the non-planarity of the interface gives rise to the instability.
 
\begin{figure*}%[h]
	\begin{center}
		\includegraphics[width=0.6\textwidth]{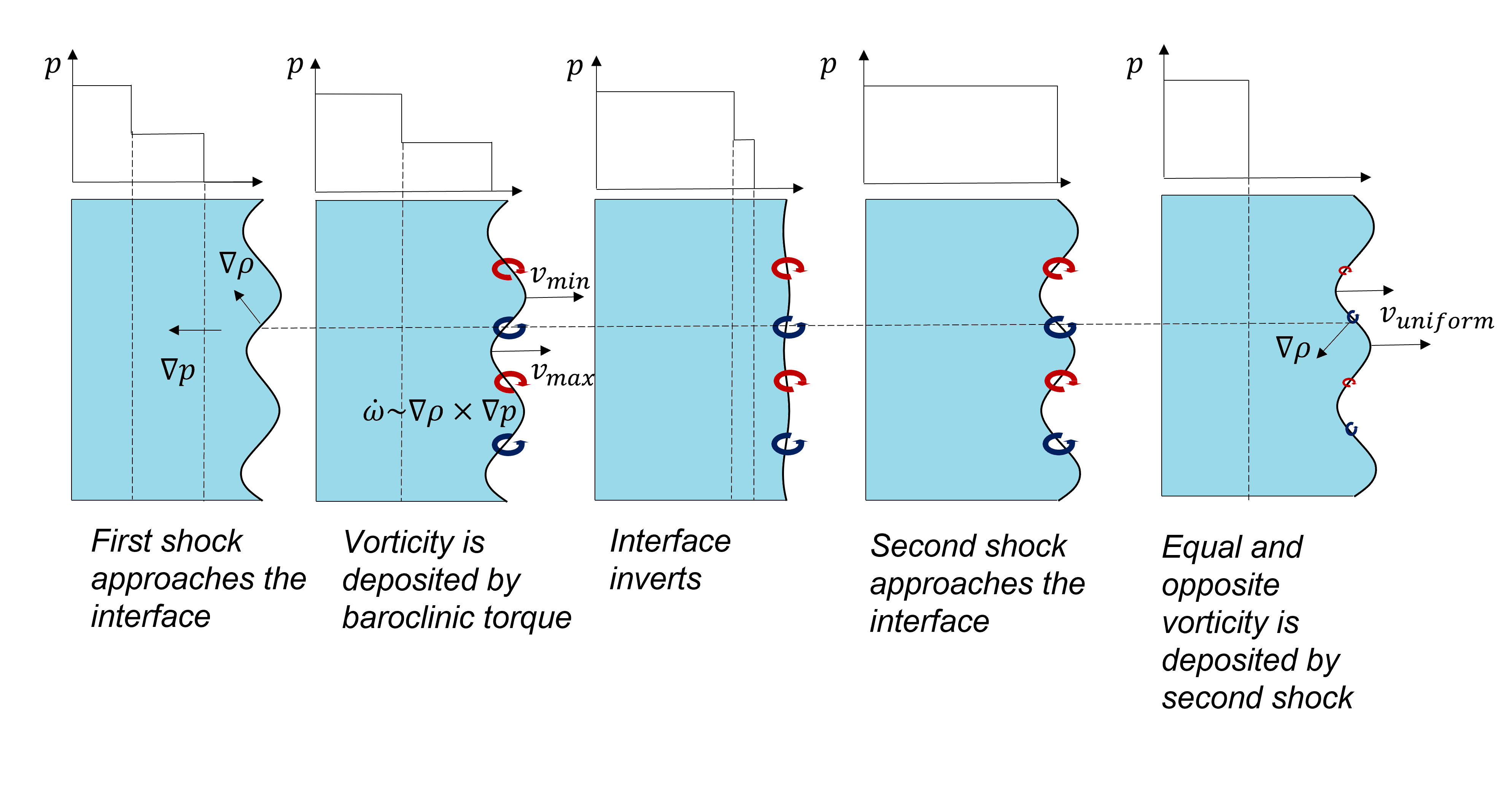}
		\caption{\small An illustration of the proposed mechanism of suppression of RMI via shock timing. A double shock wave (illustrated in gray) propagates depositing vorticity at the interface with a specific timing wherein the second vorticity deposition cancels the first due to the inversion of the wave profile.} \label{checkpoint}
	\end{center}
\end{figure*}

 In this article, we advance the following conjecture: for a heavy-light interface loaded by two shocks in sequence, owing to the fact that the interface shape inverts (i.e. valleys become peaks and peaks become valleys), there will be a special time delay between the shocks such that the vorticity deposited by the second shock will be nearly equal and opposite the vorticity deposited by the first shock; thereby canceling (potentially to zero) and leaving the interface stable. The proposed RMI suppression principle which we refer to as \textit{double shock} is illustrated in Fig.~\ref{checkpoint}.
Of course, virtually no interface in application is a perfect sinusoid \cite{cherne2015shock,mikaelian2005richtmyer}; we derive a formula for optimal shock timing corresponding to arbitrary groove shapes. This principle is extremely effective in suppressing RMI from differently shaped grooves.
\section{A basic model}
The classical paper by Richtmyer \cite{richtmyer1954taylor} modified earlier work due to Taylor \cite{taylor1950instability} to derive the velocity of amplitude growth for a single sinusoidal surface perturbation of wavenumber $ k $
\begin{equation}\label{equation 1}
\bar{v} = k \Delta u A^+ a_0 \ ,
\end{equation}
where $ \Delta u $ is the jump in particle velocity arising from the shock, $ A^{+}= (\rho_{\text{light}}-\rho_{\text{heavy}})/(\rho_{\text{light}}+\rho_{\text{heavy}}) $  is the Atwood number, and $ a_0$ is the initial amplitude.

We now suppose that the effects of two shocks in sequence may be superposed linearly giving the perturbation amplitude velocity
\begin{equation}\label{two shocks}
v(k) = v_{1}(k)+v_2(k) \ ,
\end{equation}
where
\begin{equation}\label{per shock}
v_{i}(k) = k \Delta u_{i} A^{+} a_{i,0}(k) \ , 
\end{equation}
where the index $ i=1,2 $ denotes the first or second shock. Note the dependence on wavenumber $ k $. We observe that under a constant velocity following the first shock, the initial amplitude for the second shock will be
\begin{equation}\label{a2}
a_{2,0}(k) = v_{1}(k) t + a_{1,0}(k) \ . 
\end{equation}

Now, we investigate whether there is a pair of shocks for which the amplitude growth is minimized. Minimizing $ \sum_{k} |v(k)|^2 $, and solving equations \eqref{two shocks},\eqref{per shock}, and \eqref{a2} we obtain an expression for the delay time between the two shocks as a function of the jumps in velocity
\begin{equation}\label{delay time}
t = -\dfrac{1}{  A^{+}} \pqty{ \dfrac{1}{\Delta u_{1}}+ \dfrac{1}{\Delta u_{2}}} c_g \ ,
\end{equation}
where $ c_g $ is a geometric prefactor accounting for the effects of multiple wavenumbers
\begin{equation}\label{geometricprefactor}
c_g = \dfrac{\sum_{k}^{\infty} k^3 a^2_{1,0}(k)}{\sum_{k}^{\infty}k^4 a^2_{1,0}(k)} \ .
\end{equation}
Equation \eqref{delay time} has the simple interpretation of the harmonic average of the particle velocity jumps with a prefactor coming solely from the Atwood number and the spectral content of the interface via $ c_g $. Evidently, for $ A^{+}<0 $, this equations yields $ t>0 $.  We mention that, in the special case of a single wavenumber, $ c_g =1/k $ and the velocity arrests completely; this phenomena was first pointed out by \cite{mikaelian1985richtmyer}, labeled `freeze-out', and a formula which is a special case of \eqref{delay time} was presented.

\begin{figure}[h]
	\begin{center}
		\includegraphics[width=0.9\linewidth]{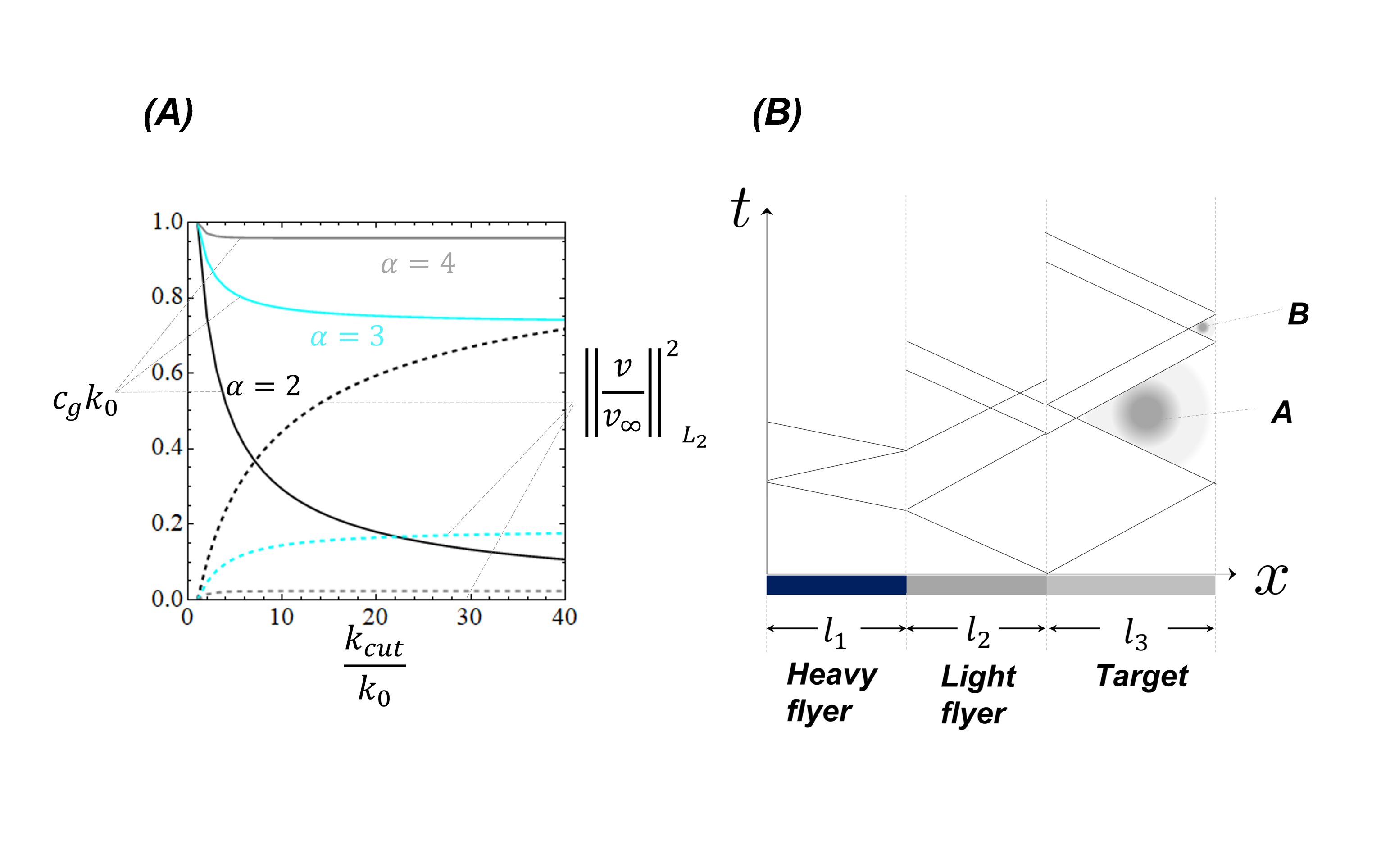}
		\caption{\small (A) We show $ c_g k_0 $ and the velocity of the suppressed jet as a function of properties of the spectral content of the original interface. (B) We sketch a notional x-t diagram for an experimental design using stepped fliers. This shows the propagation of shockwaves through the stepped flyer. We denote the particle velocities in space-time regions A and B respectively.} \label{xt}
	\end{center}
\end{figure}

Evidently, $ c_g $ depends on the spectral content of the initial interface via $ a_{1,0}(k) $. We suppose that $a_{1,0}(k) \propto k^{-\alpha}   $ for $ k \leq k_{\text{cut}} $ and $a_{1,0}(k) \propto k^{-\beta}  $ for $ k > k_{\text{cut}} $. Taking $ \alpha $ to be a small integer and $ \beta $ to be large provides a simple prototype for many realistic interfaces wherein the small scale features are much smoother below some lengthscale. Also, RMI effects of large wavenumbers may be regularized out by effects like strength, viscosity, or non-linearity.
We show in Fig.~\ref{xt}(A) the geometric factor $ c_g $ and the norm of the instability growth velocity (normalized by unmitigated jet velocity) plotted versus $ k_{\text{cut}} $ for $ \alpha = 2,3,4$. Note that we have normalized $ c_g $ and $ k_{\text{cut}} $ by the smallest wavenumber $ k_0 $. The case of $ \alpha = 2 $ is a prototype for an interface with sharp features such as a v-groove; the case of $ \alpha\geq4 $ would correspond to a highly smooth interface where the spectral content decays rapidly with increasing wavenumber. Evidently, the jetting suppression is substantial particularly for $ k_{cut} $ not too large or for $ \alpha>2 $. In light of the case of $ k_{\text{cut}}/k_0 =1 $, we observe that $ c_g k_0 $ can be interpreted as the fractional reduction in delay time between the two shock waves relative to the optimal delay time for an interface with a single wave number. %\\

To understand the shock structure, we consider the concrete case of a stepped flier. We consider the x-t diagram for the 1D setting as illustrated in Fig. ~\ref{xt} (B). The stepped flier consisting of heavy (\textit{navy}) and light (\textit{gray}) materials strikes the target (\textit{light gray}). We remark that this is, in fact, quite a bit simpler than the 2D RMI picture we have thus far introduced however we can obtain a surprising amount of insight regarding the shock strengths and timing. The classical jump conditions from the balance of momentum (neglecting dissipative effects) is
%\begin{equation}\label{jump}
$[|p|] - c_s  [| \rho u |] = 0 \ , $
%\end{equation}
where $ p $ is the pressure, $ \rho$ is the density, and $ u $ is the particle velocity. %The shock speed $ c_s $ is a function of material state and is often given $ c_s = c_0 + s v $ (e.g. \cite{meyers1994dynamic}) where $ s,c_0 $ are material specific constants.
For simplicity, we suppose that at the particle velocities experienced by the target and first flier, the shock speed $ c_s $ is nearly constant. Further, we assume that the changes in density are not too large. With these assumptions, it is easy to develop explicit expressions for the particle velocity following the arrival of the first and second shock. The particle velocity of the first $u_A$  and second shock $u_B$, respectively, are given by
\begin{equation}\label{va}
u_A = \frac{2 \rho_2  c_{s2} }{c_{s3}\rho_3+\rho_2 c_{s2} }u_0 \ , 
\end{equation}
and

\begin{equation}\label{vb}
\dfrac{u_B}{u_A}= \\
\frac{\left(3 \ c_{s3} c_{s1} \rho_3 \rho_1-c_{s3} \rho_2 \rho_3 c_{s2} +c_{s1} \rho_2 \rho_{1} c_{s2} +\rho_2^2 c_{s2} ^2\right)}{(c_{s3} \rho_3+\rho_2 c_{s2} ) (c_{s1} \rho_1+\rho_2 c_{s2} )}  \ ,
\end{equation}
%\begin{equation}\label{vb}
%\begin{split}
%&\dfrac{u_B}{u_A}= \\
% &\frac{\left(3 \ c_{s3} c_{s1} \rho_3 \rho_1-c_{s3} \rho_2 \rho_3 c_{s2} +c_{s1} \rho_2 \rho_{1} c_{s2} %+\rho_2^2 c_{s2} ^2\right)}{(c_{s3} \rho_3+\rho_2 c_{s2} ) (c_{s1} \rho_1+\rho_2 c_{s2} )}  \ ,
%\end{split}
%\end{equation}
where $ \rho_{i} $ and $ c_{si} $ are the density and shock speed of the $ i $th material indexed from left to right in Fig.~\ref{xt}.
If the constant shock speed or small density change assumption is omitted, this still may be (analytically) solvable in certain cases, however the expressions become significantly more complex. An additional complexity here is that the topology of the intersecting shock waves in Fig.~\ref{xt} may change depending on thicknesses and shock speeds; thus, the specific expression may actually change. Importantly though, the behavior of the stepped release velocity is somewhat robust with respect to these kinds of changes. %Thus, given choices of materials, the post-shock velocities are fixed as a function of density and shock speed.

We estimate the time delay between the first and second shocks as $ \tau = 2 l_{2}/c_{s2}  $ where $ l_2 $ is the thickness of the first-shock flier.
Equating this time delay and \eqref{delay time}, we obtain an expression for the thickness $ l_{2} $ which will best suppress the RMI.

%
%\subsubsection{The case of phase transitions}
%Of course, there are many ways to generate multiple shocks including other impactor driven systems, high explosive driven systems, laser or pulsed power machines. Most of these follow a similar development to the previous and we will not develop them here. We will however address the special case of phase transitions.
% 
%We now consider the behavior of a material exhibiting a phase transition. 

We now consider scenarios in which shock-induced phase transitions can suppress RMI. In such a case, the material has multiple wave-speeds; the resulting behavior can be complex and difficult to treat analytically. 
To make progress, we suppose that there is a single wavespeed in both the parent and daughter materials. 
To have a stable multi-wave structure, we must have the condition that the lower pressure phase has a faster wave speed than the higher pressure phase: $ c_{1}^{2}\geq c_2^2 \ . $
Given a certain sample size $ l $, it is easy to see that the time delay between the arrival of the first and second waves is $ \tau = l \pqty{c^{-1}_{1} - c^{-1}_{2}} \ . $ Equating this with \eqref{delay time}, we have a relationship between the length of the phase transition material and the wavenumber of the interface.

Similar to the preceding stepped flier discussion, we have the following simple estimate of the release velocity following the arrival of the first and second part of the shock wave: $u_{\text{r},1} = 2 p_{\text{tr}}/(c_{i,1} \rho_i) \ , $ and
\begin{equation}\label{ur2}
 u_{\text{r},2} =\xi ( 2 c_{w} u_{\text{imp}} \rho_{w} + \big(\dfrac{c_{w}\rho_{w}}{c_{i,2}\rho_{i}}+\dfrac{c_{i,2}}{c_{i,1}}-1\big) p_{\text{tr}} + c_{w}\rho_{w} u_{\text{r},1}  ) \ , 
\end{equation}
%\begin{widetext}
%	\begin{equation}\label{release2}
%u_{\text{release},2} = \frac{2 c_{i,1} c_{i,2} c_w \rho _i u_{\text{imp}} \rho_w-c_{i,1} c_{i,2} \rho _i p_{\text{tr}}+c_{i,2}^2 \rho _i p_{\text{tr}}+c_{i,1} c_w p_{\text{tr}} \rho_w-c_{i,2} c_w p_{\text{tr}} \rho_w}{c_{i,1} c_{i,2} \rho_i \left(c_{i,2} \rho_i+c_w \rho_w\right)}\ , 
%	\end{equation}
%\end{widetext}
where $ p_{\text{tr}} $ is the phase transition pressure, $c_{i,1} $, $ c_{i,2} $, are the shock speeds of the parent and daughter materials in the impacted material, $ c_w $ is the shock speed in the impactor material, and $\xi = (c_{i,2}\rho_{i} + c_w \rho_{w})^{-1}$ is the average shock impedance. The densities (again assumed to have small variation) are $ \rho_{w}$ in the impactor material and $ \rho_{i} $ in the impacted material.

The pressure following passage of the second shock is 
\begin{equation}\label{key}
p_2 = c_w \rho_w \xi \left( \left(c_{i,2} \rho_i u_{\text{imp}}+p_{\text{tr}}\right)-c_{i,2} p_{\text{tr}}/c_{i,1}\right) \ .
\end{equation}
This is an important term to estimate because most phase transition materials have a relatively narrow pressure range in which the double shock structure holds; if the material is overdriven then the high pressure phase shock speed will overtake the low pressure shock.
\\
An important corollary of this line of thought is that, in specific instances, a combination of experimental geometry and properties of the phase transition will lead to the self-suppression of RMI of a specific wavelength.
We remark that under very rapid compression conditions encountered in shock-compression experiments, the apparent transition pressure $p_{tr}$ can vary to some degree due to effects of kinetic processes; we do not consider such effects herein but suggest that their future study in this context may also be interesting.

\section{Simulations and Experiments}
We now present computational and experimental results which support the proposed RMI suppression strategies.

\begin{figure}[h] %[!htb]
	\begin{center}
		\includegraphics[width=0.95\linewidth]{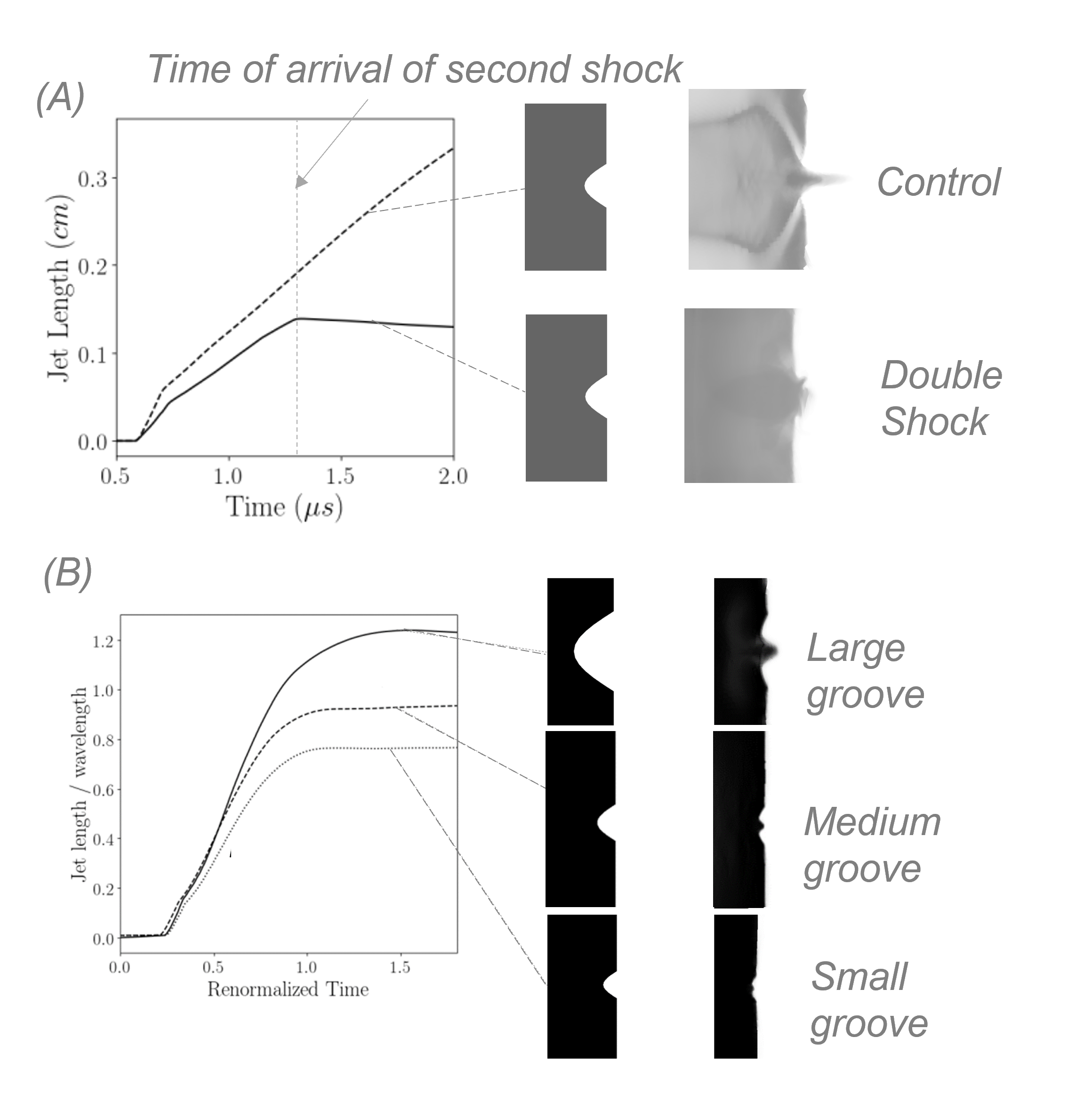}
		\caption{\small Simulations are shown for (A) a stepped flyer and (B) an iron target driven by shock loading through the phase transition. In (A), the \textit{solid} lines are from the double shock and the \textit{dashed} lines are from the control case with a single shock selected to drive the target to an equivalent particle velocity.  The asymptotic velocity of the planar free surface is virtually identical while in the double shock case, the deviation from planarity due to interface evolution is dramatically reduced. In (B), the groove size in an iron target is varied and the small groove is well-matched to the resulting double shock; evidently, the jet is minimized in this case. } \label{simulations}
	\end{center}
\end{figure}

For our simulations, we use the high order production multiphysics code MARBL \cite{dobrev2012high,dobrev2011curvilinear,anderson2018high,anderson2015monotonicity}. We show in Fig.~\ref{simulations} (A) the results of two simulations wherein the jet length is computed as a function of time; the first is driven by a double shock well matched by our theory to the groove size and the second is a control case with a single shock selected to drive a free flat surface to the same particle velocity as the double shock case.
Evidently, the suppression is very strong with the jet length arresting following passage of the second shock in contrast to a standard shock loading where the jet continues to grow linearly. 

In Fig.~\ref{simulations} (B), we illustrate the variability of jetting with groove size for a shock wave passing through an iron target; iron undergoes a phase transition from $ \alpha $ (BCC) phase to $ \epsilon $ (HCP) phase at a pressure of about $ 13.8 $ GPa which exhibits a large volume change yielding the double shock-wave structure in this case.
The analysis of instability suppression is complicated by the fact that the strength of iron is substantial; the yield strength itself will cause the jet to arrest. 
We must seek metrics which control for the sensitivity of arrest length to yield strength and demonstrate the sensitivity to double shock. 
As shown in \cite{piriz2008richtmyer}, in the absence of a phase transition, one expects the product of asymptotic jet length times wave-number to exhibit this property. Thus, we plot jet length per wave length versus time and examine the variation in the asymptotic limits. We mention that the strain rate will affect the behavior of the jetting as measured by this metric; however, given the variation in groove sizes considered here, there is unlikely to be a substantial effect.
The arrest length to wave length ratio decreases substantially as the wavelength is brought into alignment with the drive-timing provided by the experimental design.
This demonstrates that the double shock principle, in the context of phase transitions, suppresses jetting. 

We now report the results of two gas-gun experiments supporting the efficacy of the double shock principle. These experiments were conducted at Special Technologies Laboratory on their single-stage gas gun.
The basic fact that we wish to exploit is that for a given two shock structure the amplitude growth rate should depend, according to our simple theory, on the wave number. 
Thus, we specifically design dynamic experiments with two grooves, one matching the drive and one significantly different where the jetting structure is preserved.

For experiment 1, we use a 39 mm diameter stepped flier (4mm thick tantalum back plate and 2.6 mm thick PMMA front) launched at 2.3 mm/$ \mathrm{\mu s} $. The target is a 2mm thick PMMA with a 0.6 mm deep 90 degree groove and a 2.4 mm deep 90 degree groove.
Experiment 2 used a 38 mm diameter 8mm thick aluminum flier at = 1.315 mm/$\mathrm{\mu s}$ striking a $6$ mm iron target with 0.6 and 2.4 mm deep 90 degree grooves. 
In both cases, the experiments are designed so that jetting from the smaller groove should be suppressed and jetting from the larger groove will not be well suppressed.
We have used the equation of state LEOS 5060 for the PMMA target and SESAME-2140 for the iron target.
%
%We remark that there are quantitative differences between the classical sinusoidal groove and a v-groove; however, as shown in \cite{mikaelian2005richtmyer}, those effects tend to be small. 
%
%We have found in our own studies of RMI that the manufacturing of a v-groove greatly simplifies many practical aspects of fielding an experiment and that the qualitative properties which we wish to exhibit are preserved.

\begin{figure*}[h] %[!htb]
	\begin{center}
 		\includegraphics[width=1.0\linewidth]{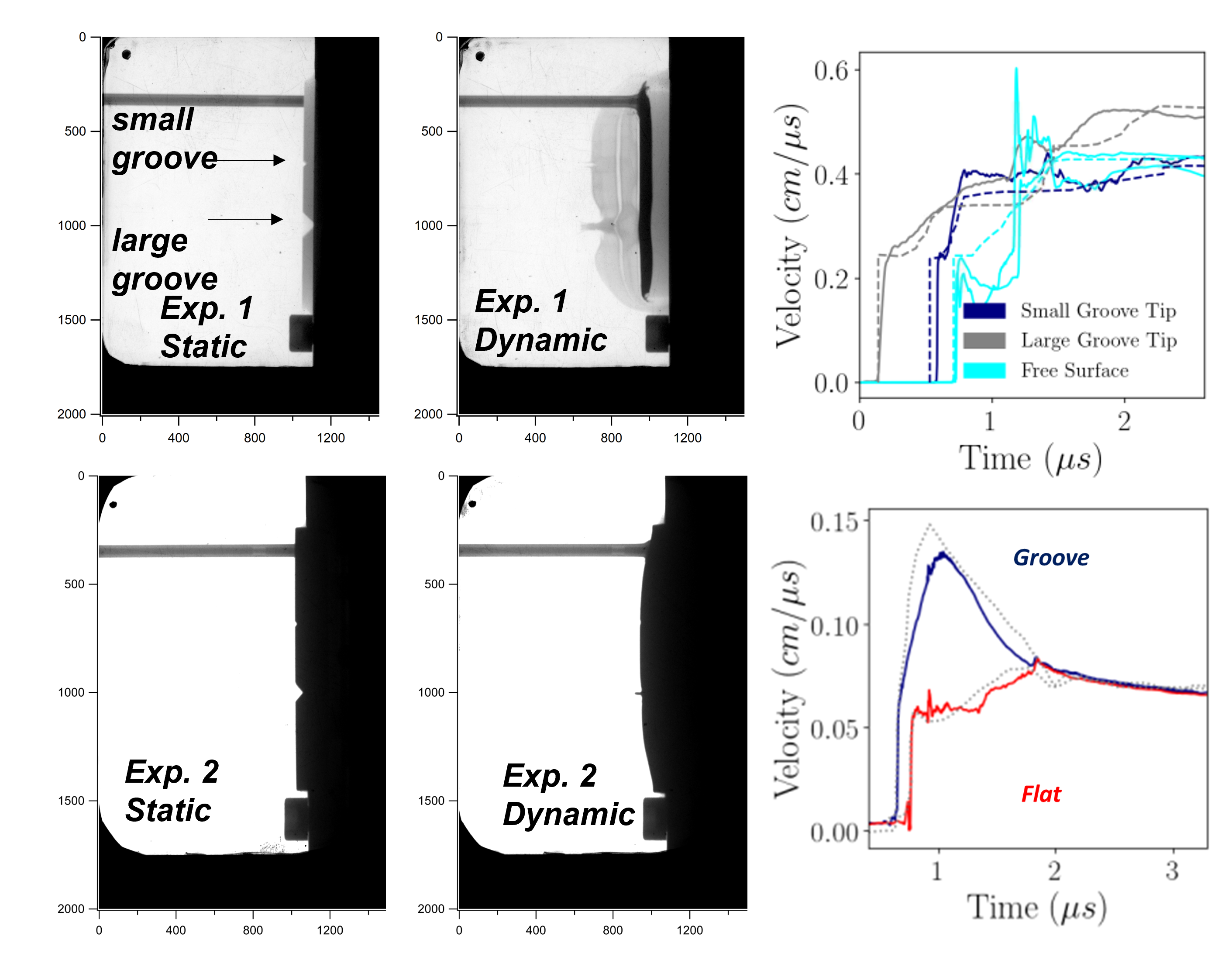}
		\caption{\small Experimental radiographs for an experiment examining experiment 1 (\textit{top left}), a stepped flyer, and for experiment 2 (\textit{bottom left}), a phase transition in iron. We compare simulated velocity traces (solid lines) directly to experimental velocimetry data (dashed lines) for experiment 1 (\textit{top right}) and experiment 2 (\textit{bottom right}). X-ray images are taken 5 $ \mathrm{\mu s} $ following impact. The experimental velocimetry measurements are shown in dashed lines and the simulated velocity histories are shown in solid lines. The jet suppression is substantial for grooves which are well matched to the double shock. We remark that there are two simulation free surface tracers given in cyan to give a sense of velocity variability along the free surface. } \label{experiments}
	\end{center}
\end{figure*}

In Fig.~\ref{experiments}, we show radiographs of the static configuration of the two experiments as well as a dynamic radiograph in late time as well as the velocimetry data for the two experiments.
In the upper left of Fig.~\ref{experiments}, the double shock experiment shows that the smaller groove (the groove well-matched to be suppressed) exhibits a strongly suppressed jet while the larger groove exhibits a standard jet. 
We remark that the smaller groove exhibits small secondary jets on the sides of the original grooves; we have observed that this occurs when the second shock arrives a little bit after optimal timing.
In the lower left of Fig.~\ref{experiments}, we show the late time behavior of the phase transformation experiment 2. 
Evidently, the jetting from the smaller groove is completely suppressed whereas the larger groove still exhibits some jetting.

On the right side of Fig.~\ref{experiments}, we show comparisons of the experimental velocimetry traces to the simulation for experiment 1 and 2 respectively. The simulations agree well with the experimental velocimetry with very simple models and no tuning of material specific parameters. In particular, we point out that the late time velocities for the small groove and the flat surface are nearly identical while the large groove velocity remains much higher. This demonstrates suppression of jetting from the smaller groove as, in the absence of a well-tuned double shock drive, the velocities of the two groove measurements should be nearly identical.

The results presented in this letter suggest that the double shock jetting suppression principle can be successfully applied in practice. In future studies, we suggest that full-field radio-graphic measurements should be made to examine in detail complex, non-sinusoidal interfaces; we suspect that a generalization to large numbers of shocks is possible and that this approach could be applied to jetting suppression in ICF.  
%In closing, we would like to emphasize a number of open questions in this line of inquiry: How do we find an in-depth understanding of the evolution of strongly non-sinusoidal interfaces and do these deviate from the idealized case? Does the double shock mechanism generalize to multi-shock sequences in a natural way? Can this suppression mechanism be utilized to improve the performance of ICF at the NIF? 

%\Floatbarrier

\begin{acknowledgments}
This work was performed under the auspices of the U.S. Department of Energy by Lawrence Livermore National Laboratory under Contract DE-AC52-07NA27344 and was supported by the LLNL-LDRD Program under Project No. 21-SI-006. Lawrence Livermore National Security, LLNL-JRNL-846621. 
\end{acknowledgments}

%\appendix

%\section{Appendixes}

%To start the appendixes, use the \verb+\appendix+ command.
%This signals that all following section commands refer to appendixes
%instead of regular sections. Therefore, the \verb+\appendix+ command
%should be used only once---to setup the section commands to act as
%appendixes. Thereafter normal section commands are used. The heading
%%for a section can be left empty. For example,
%\begin{verbatim}
%\appendix
%\section{}
%\end{verbatim}
%will produce an appendix heading that says ``APPENDIX A'' and

%In Fig.~\ref{phasetrans}, we show the materials behavior for iron, a %canonical example of a phase transformation with a relatively large %volume collapse.

%\begin{figure}[h]
%	\begin{center}
%		\includegraphics[width=0.4\linewidth]{images/Phase_transition_in%_iron}
%		\caption{\small We show the pressure and sounds speed versus %density for iron near its BCC-HCP phase transition.} %\label{phasetrans}
%	\end{center}
%\end{figure}

%\nocite{*}

\bibliography{research}
\bibliographystyle{unsrt}
\end{document}